\documentclass{emulateapj} 

\usepackage{graphics}

\shorttitle{The mass and radius of the Rapid Burster}
\shortauthors{Sala et al.}

\begin{document}

\title{Constraints on the mass and radius of the accreting neutron star in the Rapid Burster}

\author{G. Sala\altaffilmark{1,2}, F. Haberl\altaffilmark{3}, J. Jos\'e\altaffilmark{1,2}, A. Parikh\altaffilmark{1,2}, 
 R. Longland \altaffilmark{1,2}, L.C. Pardo\altaffilmark{4}, M. Andersen \altaffilmark{5}}

\altaffiltext{1}{Departament de F\'isica i Enginyeria Nuclear, EUETIB,
Universitat Polit\`ecnica de Catalunya, c/ Comte d'Urgell 187, 08036 Barcelona, Spain}
\altaffiltext{2}{Institut d'Estudis Espacials de Catalunya, c/Gran Capit\`a 2-4, 
Ed. Nexus-201, 08034, Barcelona, Spain}

\altaffiltext{3}{Max-Planck-Institut f\"ur extraterrestrische Physik,
Giessenbachstra{\ss}e, D-85748, Garching, Germany}

\altaffiltext{4}{Grup de Caracteritzaci\'{o} de Materials,
Departament de F\'{\i}sica i Enginyieria Nuclear, ETSEIB,
Universitat Polit\`ecnica de Catalunya, Diagonal 647, 08028
Barcelona, Catalonia, Spain
}

\altaffiltext{5}{Research \& Scientific Support Department,    European Space Agency,    ESTEC,    Keplerlaan 1,    2200 AG Noordwijk,    Netherlands}

\begin{abstract}

The Rapid Burster (MXB 1730-335) is a unique object, showing both type I 
and type II X-ray bursts. 
A type I burst of the Rapid Burster was observed with Swift/XRT on 2009 March 5,
showing photospheric radius expansion for the first time in this source. 
We report here on the mass and radius determination from this 
photospheric radius expansion burst using a Bayesian approach.
After marginalization over the likely distance of the system (5.8--10 kpc) 
we obtain M=1.1$\pm$0.3~M$_{\odot}$ and R=9.6$\pm$1.5~km 
($1\sigma$ uncertainties) for 
the compact object, ruling out the stiffest equations of state for the neutron star.
We study the sensitivity of the results to the distance, the color correction factor, and 
the hydrogen mass fraction in the envelope. We find that only the distance plays a crucial role.

\end{abstract}


\section{Introduction}

X-ray bursts were discovered in 1975 by \cite{gri76} in observations
performed with the Astronomical Netherlands Satellite of a previously 
known X-ray source, 4U 1820-30. Similar episodes had been observed
in 1969 in the Norma constellation, with two Vela satellites \citep{bel72}, 
but the observed feature was related to an accretion event and 
not identified as a new type of source until 1976 (Belian et al. 1976; 
see also Kuulkers et al. 2009 for a reanalysis of that event).
The first discovered burst (in Cen X-4) is still the nearest-known 
X-ray burst source (at 1~kpc) and the brightest X-ray burst ever recorded 
($\sim$50~Crab \footnote{1 Crab = 2.4$\times10^{−8}$ erg s$^{-1}$ cm$^{-2}$ in the 2--10 keV band}).
These pioneering discoveries were soon followed by the identification of 
three additional bursters, one of which being the enigmatic Rapid Burster 
(MXB 1730-335, Lewin 1976). 
To date, more than 90 Galactic X-ray burst sources have been identified \citep{liu}.

\cite{vanhorn74} and \cite{han75} were the first to suggest that 
nuclear burning on the surface of neutron stars could be unstable.
Once X-ray burst sources were successfully identified in nature, 
Maraschi \& Cavaliere (1976), and independently, Woosley \& Taam (1976), 
proposed that X-ray bursters could be powered by thermonuclear runaways on the surface of 
accreting neutron stars. However, it was soon realized that the quick 
succession of flashes exhibited by the Rapid Burster (with recurrent 
times as short as 10 sec), did not match the general pattern shown 
by the other bursting sources. A classification of type I and type II bursts was 
then established \citep{hof78}, the former associated with thermonuclear flashes, 
the later linked to accretion instabilities.

The Rapid Burster is located in the Galactic bulge globular cluster Liller 1  \citep{Lewin1976}, 
at a distance between 5.8 and 10 kpc from the Earth \citep{Ortolani2007}. 
A radio transient counterpart was detected \citep{moore}
but no IR or optical counterpart has been identified so far \citep{homer}.
It is one of the only
two known sources showing type II bursts, together with the Bursting Pulsar (GRO 1744-28),
but the Rapid Burster is unique in that it exhibits 
both type I and type II X-ray bursts (for a review see Lewin et al. 1995).
Type I (thermonuclear) bursts display
a spectral softening during the burst tail, reflecting a decrease in the effective temperature and thus the cooling 
of the neutron star atmosphere. In contrast,
type II bursts exhibit a constant temperature for the duration of the burst, which can range from $\sim$2 s to $\sim$700 s, with intervals between bursts from $\sim$7 s to $\sim$1 hr.
Over long time-scales, as daily monitored by the All Sky Monitor on board the Rossi X-ray Time Explorer, the Rapid Burster usually radiates at low X-ray flux 
($<$ 10 mCrab) indicating a low mass accretion rate. Every 100-200 days it enters into an outburst
\footnote{Note that in the X-ray binary context, outbursts refer to a period of 
higher flux due to an increased accretion rate, not to be confused with 
explosive, thermonuclear events like nova outbursts.}, reflected in a higher persistent emission
that indicates a higher mass accretion rate, 
frequent type II X-ray bursts, and some type I X-ray bursts.
Outbursts start with a fast rise up to 100 mCrab, 
followed by a decay over two to four weeks.
We triggered Swift Target of Opportunity
observations to follow the bursting
behavior during the January-March 2009 outburst. Type II bursts are
present in all our observations, 
but their analysis is not the scope of the present paper.

Here we report on one type I (thermonuclear) burst on 2009 March 5 with
expansion of the photosphere (as shown in section 3), which allows us to constrain the 
mass and radius of the neutron star in the Rapid Burster.
\cite{gal} analyzed all 66 type I X-ray bursts of the 
Rapid Burster available in the public archive of RXTE, 
and found none exhibiting photospheric expansion.
The event analyzed in the present work is thus the first one 
that makes it possible to constrain the mass and radius of the neutron star in the Rapid Burster.
In section 2 we present the observations and the analysis methods; section 3 shows the 
spectral analysis of the photospheric radius expansion event (PRE); the determination of the 
mass and radius of the neutron star with their uncertainties is presented in section 4;
we discuss the results and study their sensitivity to the distance, color correction factor and
hydrogen mass fraction in section 5; section 6 contains a summary and the conclusions.

\section{Observations and data reduction}

Daily Swift X-ray Telescope (XRT, Burrows et al 2005) monitoring 
observations were performed between 
February 26 and March 5, accumulating more than 20~ks of data \citep{Sala2009}. 
All observations show intense type II bursting activity, 
with burst intervals between 25 and 200 seconds. 
Due to the high average flux of the source, all XRT observations were taken in 
Window Timing (WT) mode, which 
minimizes the pile-up at the brightest peaks of the bursts. 
In WT mode data are free of pile-up for count-rates below 100 cts/s.
As explained below, our results are based on time resolved spectral analysis, 
with the two spectra of interest having count-rates of 62 and 25 cts/s, well 
below the threshold for pile-up in WT mode.
Swift XRT software was used for data reduction (xrtpipeline) and the HEASOFT tool xselect for 
light-curve and spectra extraction. For each of the spectra, individual exposure maps and 
ancillary files were created, and the XRT response matrix v013 was used for all 
spectral fits. XSPEC 12.6 was used for spectral analysis. 
Due to the small number of counts in our spectra, we use the C-statistic \citep{cash} in all fits.

Most of the detected type II bursts  
peak at more than 100~cts/s in the first Swift/XRT (WT) sequences (Feb. 26), which implies 
fluxes of about $10^{-8}$~erg s$^{-1}$ cm$^{-2}$ in the range 1--10~keV assuming blackbody emission
(L$_{peak}=(1-1.6)\times10^{38}$~erg/s at 10 kpc). 
The peak intensity decreases in the last Swift observations, with most bursts 
showing maximum intensities of less than 60~cts/s in XRT (WT) on 2009 March 5.

\begin{figure}
\bigskip
\epsscale{1.}
\plotone{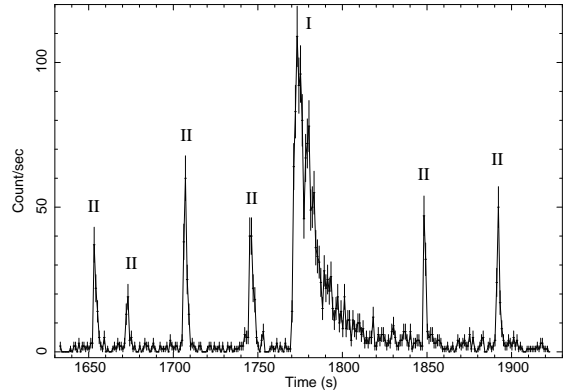}
\caption{Swift/XRT light curve (0.2--10keV) on 2009 March 5, 
during the observation sequence containing the PRE burst. Six type II burst 
can be seen in addition to the PRE type I burst. The time bin is 1s, the start time 
of the observation is 14895 at 13:27:13.216 in truncated Julian Days (TJD, Julian Day - 2440000.5).}
\label{fig:lc}
\end{figure}

\begin{figure}
\bigskip
\epsscale{1.}
\plotone{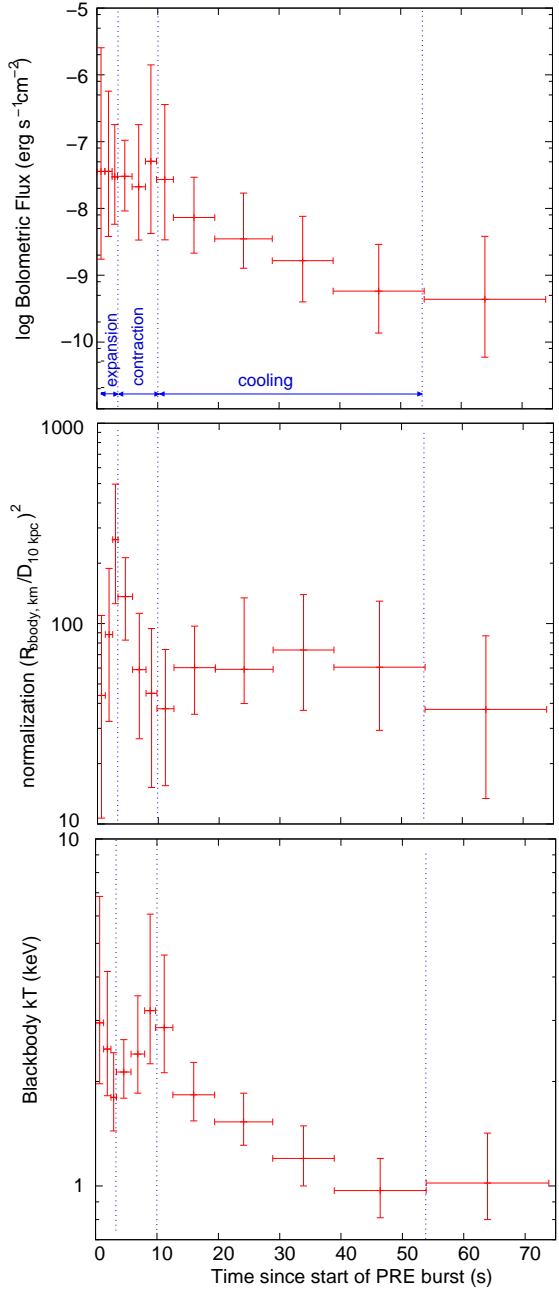}
\caption{Time resolved spectral analysis: evolution of the bolometric flux and the 
spectral parameters during the PRE burst.}
\label{fig:evol}
\end{figure}

Figure \ref{fig:lc} shows the Swift/XRT light-curve of the Rapid Burster 
during the first observation sequence in 2009 March 5 (OBS ID: 00031360008).
A bright type I X-ray burst is observed among six other fainter, type II bursts. 
For time resolved spectral analysis of the cooling tail of this type I burst, 
spectra have been extracted for a total of 12 time bins,
with a variable duration depending on the rate, ranging from 0.9 to 20 seconds (Figure \ref{fig:evol}).
The time bins have been carefully selected to follow the variations of flux during the burst while keeping the 
maximum possible statistical quality. 

For each time bin, a source spectrum is extracted, selecting the events in a circular region with a radius of 
20 pixels ($\sim$47 arcsec) centered at the source position. 
It is common in burst studies to subtract off the continuum, 
persistent emission from the burst spectrum, 
but this is known to be a source of potential errors \citep{van1986}. 
In our case, the persistent emission during the orbit containing the type I burst 
(corresponding to all time intervals free of type II bursts) 
is at least two orders of magnitude fainter than the burst spectrum, even for the faintest
time bins considered for the mass and radius determination in Section 4. 
To check the possible effect of the persistent emission in our results, 
we performed the spectral analysis for the faintest spectra
of the type I burst with and without persistent emission subtraction. 
The resulting spectral fit parameters 
are identical within less than 1\%, much smaller than our final uncertainties,
confirming that the persistent emission does not affect our results.
The systematic uncertainty in the Swift/XRT absolute calibration, considered to be 10\%
\citep{Godet}, is also negligible compared to our final 
uncertainties and does not affect our results.

\begin{figure}
\bigskip
\epsscale{1.}
\plotone{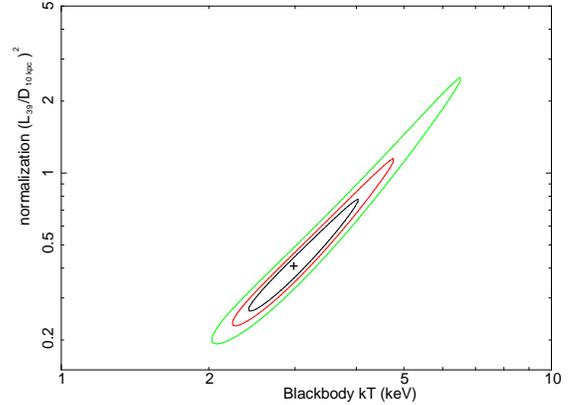}
\caption{Confidence contours at 68\%, 90\% and 99\% confidence level for the 
spectral parameters at touchdown. Normalization is in units of $L_{39}/D^2_{10kpc}$, with
$L_{39}$ being the bolometric luminosity in units of $10^{39}$ erg/s and $D_{10kpc}$ the 
distance in units of 10 kpc.}
\label{fig:td}
\end{figure}

\begin{figure}
\bigskip
\epsscale{1.}
\plotone{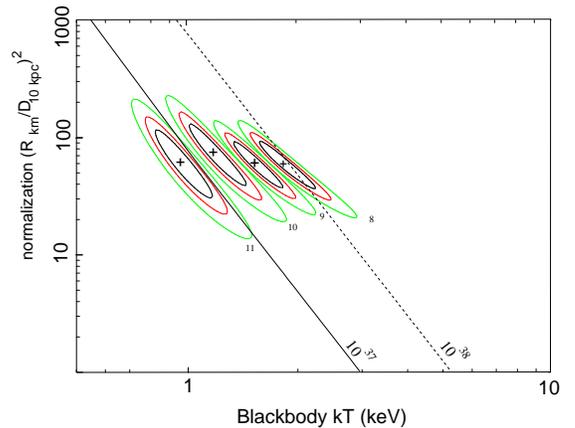}
\caption{Confidence contours at 68\%, 90\% and 99\% confidence level for the 
spectral fit parameters during the cooling tail. Normalization is in units of $(R_{km}^{bbody}/D_{10kpc})^{2}$, where $R_{km}^{bbody}$ is the apparent radius in km and $D_{10kpc}$ the distance to the source in units of 10 kpc.
Numbers by the contours indicate
the time bin (as shown in Fig \ref{fig:evol}). Diagonal lines corresponding to 
constant luminosity are shown for luminosities $10^{37}$ and $10^{38}$~erg/s.}
\label{fig:cool}
\end{figure}

\section{Spectral analysis}

All burst spectra were fit in the energy range 1--10 keV with a blackbody model, modified with 
interstellar absorption by the T\"ubingen-Boulder absorption model \citep{wil00}.  
This model is used to calculate the cross section for X-ray absorption as the sum of the cross 
sections due to the gas-phase, the grain-phase, and the molecules in the interstellar medium. 
The hydrogen density column is kept constant to its average interstellar value, $1.7\times10^{22}$cm$^{-2}$ \citep{DL}.
Note that using the relation of \cite{predehl}, this column density
corresponds to an extinction of E(B-V)=3, as observed for the Liller 1 globular 
cluster \citep{barbuy}. \cite{marshall} found the same absorbing hydrogen column density
to the Rapid Burster with Chandra/HETG spectra of type II X-ray bursts, 
indicating that there is no source intrinsic absorption.
The blackbody model used for the fit of the cooling tail spectrum
is parametrized with the effective temperature and a normalization 
constant proportional to the emitting area, more precisely 
$A_{\infty}=(R_{km}^{bbody}/D_{10kpc})^{2}$ (with $R_{km}^{bbody}$ being the radius in km
and $D_{10kpc}$ the distance to the source in units of 10~kpc).
The evolution of the bolometric flux, the normalization parameter A$_{\infty}$, and the blackbody temperature
obtained from the spectral fits is shown in Figure \ref{fig:evol}.

Three clear phases are visible in the type I burst light curve and spectral parameters evolution,
clearly indicating that photospheric radius expansion (PRE) is occurring.
First, the expansion of the photosphere occurs during the initial rise in flux, 
lasting $\sim 3$ seconds. 
Second, after maximum expansion, the contraction of the photosphere follows, 
lasting $\sim 7$ seconds. The contraction finishes at the so called 
touchdown of the photosphere (occurring at bins 6-7 in  Figure \ref{fig:evol}, 8-12 seconds after the 
start of the burst). At this moment, the photospheric radius returns to the neutron
star surface and the blackbody radius reaches its minimum value and the color temperature its maximum.
Finally, the cooling down of the envelope is observed (time bins 8-11, 13-54 seconds after the start of the burst). 
The twelfth and last bin analyzed and shown in 
Figure \ref{fig:evol} does clearly show that neither the flux nor the temperature are decreasing anymore, 
so we consider the cooling tail to finish 54 seconds after the start of the burst. 

\begin{figure}
\bigskip
\epsscale{1.}
\includegraphics[totalheight=8cm, angle=-90]{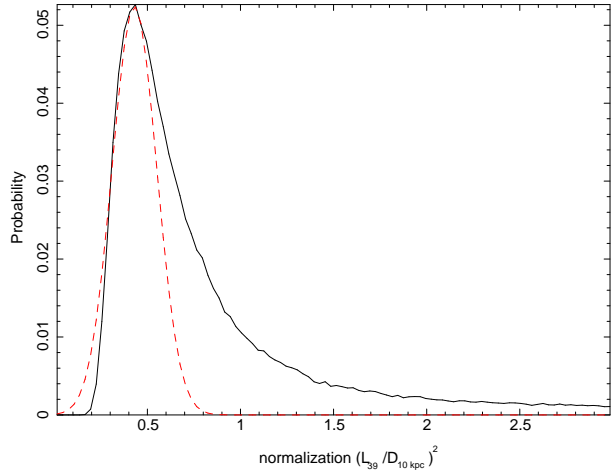}
\caption{Probability density function of the touchdown flux. Normalization is in units of $L_{39}/D^2_{10kpc}$, where $L_{39}$ is the bolometric luminosity in units of $10^{39}$~erg/s.
A Gaussian is shown in red, dashed line for comparison.}
\label{fig:pf}
\end{figure}

\begin{figure}
\bigskip
\epsscale{1.}
\includegraphics[totalheight=8cm, angle=-90]{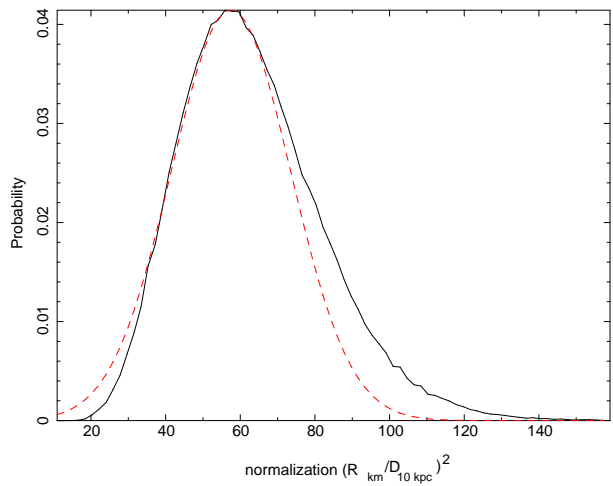}
\caption{Probability density function of the cooling area. Normalization units are $(R_{km}^{bbody}/D_{10kpc})^2$. 
A Gaussian is shown in red, dashed line for comparison.}
\label{fig:pa}
\end{figure}

We devote some attention to quantify the radius expansion. 
A straightforward calculation of the $\chi^{2}$ value with respect to the average radius would not be correct, 
because due to the low statistics of the data, the probability density function (PDF) of the spectral parameters 
in each of the time bins follows a Poisson distribution and differs clearly 
from a normal distribution. In other words, error bars are not symmetric, as shown for the 
touchdown and cooling tail bins in Figures \ref{fig:pf} and \ref{fig:pa}. 
To deal properly with the PDF of each of the bins of the radius evolution and evaluate the radius expansion,
we have taken into account the calculated PDF associated to each bin and fit each of them 
by a Poisson distribution. We have then performed a model selection using the 
Bayesian algorithm FABADA\footnote{http://gcm.upc.edu/members/luis-carlos/luis-carlos-pardo/bayesiano} 
proposed in \cite{fabada}. 
We first fit the radius evolution shown in Figure \ref{fig:evol} with a constant radius 
(more precisely, with $(R_{km}^{bbody}/D_{10kpc})^{2}=70$),
redefining $\chi^{2}$ as in \cite{fabada2}
to take into account the asymmetry of the PDF describing the data. 
The $\chi^{2}$  obtained modeling the data by a constant radius is
16.5 (for 5 degrees of freedom), with a chance probability of only 0.55\%. 
In contrast, a model consisting of an expansion and contraction of the radius 
provides $\chi^{2}$=3.5 (for 3 degrees of freedom). 
Therefore, the expansion is clearely favored in comparison to a constant radius.  

The bolometric luminosity is almost constant during the envelope
expansion (0--3 s) and contraction (3--12 s), and decreases during
the cooling tail (11--53 s).
Expansion of the photosphere is caused by the luminosity
reaching the Eddington limit. The neutron star mass can then be
determined from the observed bolometric luminosity, assuming that the touchdown
(i.e. maximum) luminosity corresponds to the Eddington limit as
seen by a distant observer. After the touchdown, the emitting area is assumed to be 
the whole neutron star surface, and from the spectral fits during the cooling tail, the
neutron star radius can be inferred. 
For convenience in the analysis, since we are interested in the 
cooling emitting area and the touchdown flux, the blackbody model used for the cooling tail  
is parametrized with the normalization proportional to the emitting area 
($A_{\infty}=(R_{km}^{bbody}/D_{10kpc})^{2}$, as explained above for all time bins),
while for the touchdown spectrum we use a normalization proportional to the 
bolometric flux as $L_{39}/D_{10kpc}^2$ , where $L_{39}$ is the bolometric luminosity 
in units of $10^{39}$~erg/s.

For the determination of the touchdown flux, we merge time-bins 6 and 7 of 
Figure \ref{fig:evol} to obtain a 
higher signal-to-noise spectrum. Figures \ref{fig:td} and \ref{fig:cool} 
show the confidence contours 
of the spectral parameters obtained for the touchdown spectrum and the 
spectra of the four time bins of the cooling tail. Since the count-rate 
is decreasing during the cooling tail,
the parameters become less well constrained for the later time-bins 
(labeled as 9, 10 and 11 in Figure \ref{fig:cool}). 
For the determination of the cooling emitting area, 
we consider the range of values for the  
$A_{\infty}=(R_{km}^{bbody}/D_{10kpc})^{2}$
obtained for the first bin of the cooling tail (bin 8), 
which has the brightest spectrum and thus the better constrained emitting area. 
Note that all later bins, while having
larger uncertainty ranges, are compatible with the results obtained from bin 8.

\section{Mass and radius of the neutron star in the Rapid Burster}

The time resolved spectral analysis of X-ray data of a PRE burst initially provides only the 
best-fit blackbody parameters. But there are a number of factors that play a role in the
determination of the actual neutron star radius and mass. 
Firstly, the neutron star spectrum is 
known to differ from a perfect blackbody, with the bursting neutron star atmosphere 
emitting a harder spectrum than a blackbody 
due to electron scattering in the hot neutron star atmosphere and the resulting 
suppression of emissivity \citep{vanparadijs}.
The actual effective temperature can be related to the color or blackbody temperature obtained
from the spectral fit by the color correction factor, $f_c=T_{bbody}/T_{eff}$. 
A number of recent works have been devoted to the
modeling of the neutron star atmosphere spectra and the determination of $f_c$
\citep{madej,majczyna, suleimanov}. 
The color correction factor  $f_c$ will affect both the temperature and the emitting area, 
since this is obtained from the observed flux and blackbody temperature.
A second factor affecting the radius and mass determination of the neutron star 
is the distance D, needed for the luminosity determination. 
The hydrogen mass fraction X plays its role in the Eddington limit 
for the luminosity through the electron scattering opacity, 
$\kappa_{e}=0.2(1+X) cm^{2}g^{-1}$. Finally,
the high surface gravity of the neutron star causes light deflection, affecting the observed flux and 
emitting area. In a PRE burst, the gravitational redshift changes with the expansion of the 
photosphere and special caution is needed in the selection of the
moment for the determination of the bolometric flux. 
The best moment for the determination of the bolometric flux is at touchdown, when the 
luminosity is still at the Eddington limit, but the photospheric radius is 
already at the neutron star surface, with the redshift then determined by 
the neutron star radius.

Combining the above considerations, the observed touchdown flux F$_{\infty}$ and the 
apparent angular emitting area A$_{\infty}$ depend on 
the actual neutron star radius R and mass M as 

\begin{equation}
F_{\infty}=\frac{GMc}{0.2(1+X)D^2} \left( 1-\frac{2GM}{Rc^2}\right)^{1/2} 
\end{equation}

and

\begin{equation}
A_{\infty}=\frac{R^2}{D^2f^4_{c}} \left( 1-\frac{2GM}{Rc^2}\right)^{-1} 
\end{equation}

With the uncertainty ranges of our observed touchdown flux and cooling emitting area 
as shown in the confidence contour plots in 
Figures \ref{fig:td} and \ref{fig:cool}, an accurate treatment of the error ranges is needed.
Here we follow the Bayesian approach introduced 
in \cite{ozel2006} and  \cite{ozel2009} in the first two of a series of papers by these authors on the 
determination of the mass and radii for several bursting neutron stars.
A probability density function (PDF) is assigned 
to each of the observable quantities (touchdown flux, cooling area, distance), designated by 
P(F$_{\infty}$)dF$_{\infty}$, P(A$_{\infty}$)dA$_{\infty}$, P(D)dD, as well as the color correction factor and the hydrogen mass fraction,
P(f$_{c}$)df$_{c}$ and P(X)dX. Since these are all independently determined parameters, 
the combined probability for given values of F$_{\infty}$, A$_{\infty}$, D, $f_{c}$ and X is directly 
the product of the individual probabilities. 
This combined PDF is then converted to the PDF of the neutron star mass and radius 
using the change of variables as given in equations (1) and (2).

We assign uniform probabilities to the distance, 
between 5.8 and 10 kpc 
(Ortolani et al. 2007, see however next section);
to the mass hydrogen fraction, between 0 and 0.7 (solar); and to the color correction factor, 
between 1.3 and 1.4. This last choice is based on the values obtained from theoretical modeling 
of bursting neutron star atmospheres \citep{madej,majczyna, suleimanov} and the 
fact that the ratio L/L$_{Edd}$ at the time we measure 
the emitting area is in the range 0.03-0.4.

For the observable parameters F$_{\infty}$ and A$_{\infty}$, instead of assigning a Gaussian distribution 
with the values of the spectral fit, we determine the PDF of the observed touchdown flux, P(F$_{\infty}$)dF$_{\infty}$, and of the 
cooling area, P(A$_{\infty}$)dA$_{\infty}$,  directly  from the blackbody model fit to the Swift/XRT 
spectra by running a Monte Carlo Markov chain (MCMC) to each of the spectral fits (touchdown and cooling). 
This choice is based on the fact that the results of the MCMC show that both 
P(A$_{\infty}$)dA$_{\infty}$ and P(F$_{\infty}$)dF$_{\infty}$ are asymmetric functions and differ considerably from a Gaussian distribution,
as shown in Figures \ref{fig:pf} and \ref{fig:pa}.

\begin{figure}
\bigskip
\epsscale{1.}
\plotone{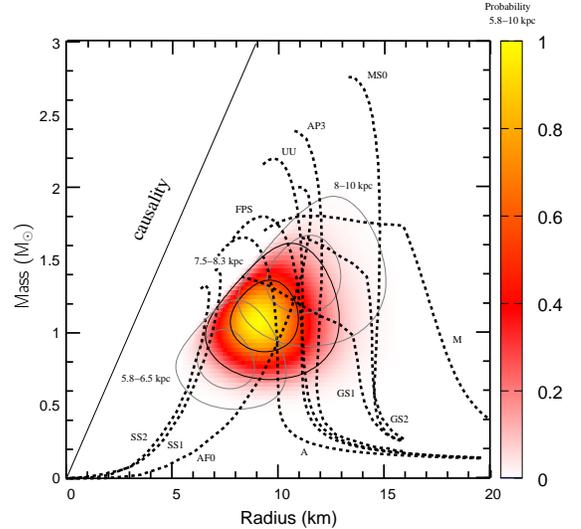}
\caption{Mass and radius of the neutron star in the Rapid Burster as determined from the 
type I X-ray burst with Photospheric Radius Expansion observed by Swift/XRT on 2009 March 5.
The color map shows the probability density function after marginalization over 
the whole range of the distance (5.8--10~kpc), 
the color correction factor (1.3--1.4) and the hydrogen mass fraction (0--0.7).
Confidence contours at 1$\sigma$ and 2$\sigma$ are shown for the distance ranges 5.8--6.5~kpc, 
7.5--8.3~kpc, and 8--10~kpc, 
corresponding to different distance determinations in Ortolani et al. (2007) and Valenti et al. (2010). 
Mass-radius relations for several equations of state are shown as well, with relations labeled
with MS0, AP3, UU, FPS and A corresponding to equations of state of neutron stars without condensates, 
M, GS1 and GS2 to equations of state with condensates, and AF0, SS1 and SS2 to strange quark matter stars 
(see Lattimer and Prakash 2001 for more details). 
The upper left part of the plane (left of the solid line) is excluded by causality.}
\label{fig:mr}
\end{figure}

To compute the final PDF on the M-R plane, we proceed as follows:
for each M-R pair we calculate the corresponding values of A$_{\infty}$ and F$_{\infty}$ for given values of 
D, f$_{c}$ and X within the range considered, with equations (1) and (2); we then
determine the P(A$_{\infty}$)dA$_{\infty}$ and P(F$_{\infty}$)dF$_{\infty}$ corresponding to these values from our observational results
and calculate the combined PDF. We marginalize over (integrate
over all values of) D, f$_{c}$ and X to obtain the probability density function on the M-R plane as
shown by the color map in Figure \ref{fig:mr}. In the next section we test the
robustness of the results to variations of the distance, color correction factor and mass hydrogen fraction.

Finally, we obtain the PDF of the neutron star mass by marginalizing over the radius, and
the PDF of the radius by marginalizing over the mass. 
The final radius PDF is well fit with a 
lognormal distribution 
with median  9.6~km and a variance of 2.5~km
(a Gaussian fit provides a similar mean 9.6 km and $\sigma$=1.5~km).
The final PDF of the mass is acceptably represented by a Gaussian distribution with mean 1.1~M$_{\odot}$ and $\sigma$=0.3~M$_{\odot}$.

\begin{figure}
\bigskip
\epsscale{1.}
\plotone{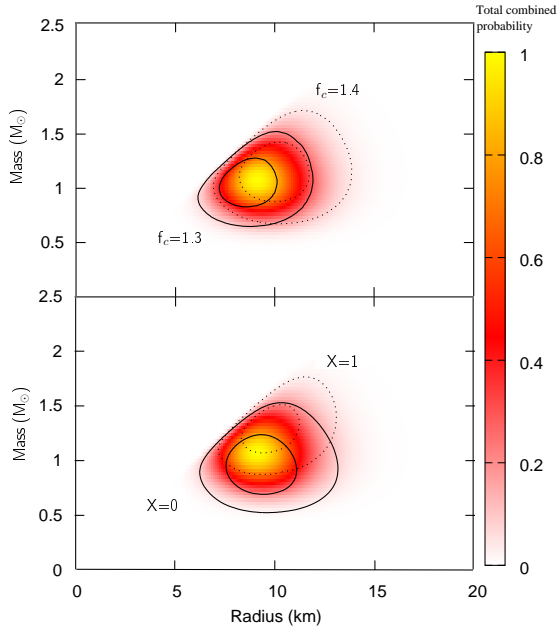}
\caption{Sensitivity of the final mass and radius of the neutron star to the  
color correction factor f$_c$ (upper panel) and the hydrogen mass fraction X
(lower panel). 
Confidence contours at 1$\sigma$ and 2$\sigma$ are shown for
f$_c$=1.3 (solid lines) and f$_c$=1.4 (dotted lines), and for 
X=0 (solid lines) and X=1 (dotted lines), while the color scale shows
the probability distribution function as in Figure \ref{fig:mr}, for 
f$_c$ marginalized over 1.3--1.4 and X over 0--0.7.}
\label{fig:x}
\end{figure}

\section{Discussion}

Time resolved spectroscopy of the type I X-ray burst from the Rapid Burster
with photospheric radius expansion allows us to 
constrain the mass and radius of the accreting neutron star 
by determining the probability density function on the mass-radius plane, 
as shown in Figure \ref{fig:mr}. For the accreting neutron star in the Rapid Burster we obtain
M=1.1$\pm$0.3~M$_{\odot}$ and R=9.6$\pm$1.5~km ($1\sigma$ uncertainties) for 
the assumed distance range 5.8--10 kpc. This indicates that 
the accreting object in the Rapid Burster is compatible with both normal matter and  
strange quark matter equations of state, but the stiffest equations of state 
can be ruled out. 

This is, to our knowledge, the first detected 
type I X-ray burst of the Rapid Burster showing
photospheric radius expansion. \cite{gal} analyzed all the observations 
of the Rapid Burster available in the RXTE archive and identified 66 type I 
X-ray bursts. While some of those bursts show a flat topped light curve, none of them
follows a temperature and radius evolution like the one shown in our Figure \ref{fig:evol}.
In addition, the peak flux of the X-ray burst analyzed in the present work 
is about a factor 3 higher than the peak flux of the brightest type I 
X-ray bursts of the Rapid Burster shown by \cite{gal}, supporting that
in the present case the Eddington limit of the luminosity was reached.

We investigate now in detail the effect of these choices and the sensitivity of the 
final result to these parameters. 
We consider first the effect of the distance on the final results.
\citet{Ortolani2007} determined the distance to Liller 1 based on HST NICMOS F110W and F160W band photometry and by using the location of the horizontal branch in the color-magnitude diagram compared to other globular clusters with known distances. 
The F110W and F160W filters are different than the ground-based equivalent J and H band filters in that they are wider and extend further to the blue than the ground-based filters.
A careful transformation of NICMOS photometry into a ground-based system is therefore necessary.
This is particularly true for observations of Liller 1, 
which suffer from some foreground extinction of E(B-V)=3.0 \citep{barbuy}.
Unfortunately, as indicated in \citet{Ortolani2007}, the choice of the color transformation is of importance, in particular for the J-H color term.

\cite{Ortolani2007} determined the distance to Liller 1 assuming metallicities  [Fe/H]=-0.1 and
 [Fe/H]=-0.6, and for three different photometric calibrations, reporting a total of
six distance estimates in the range 5.8--10.0~kpc. 
More recently, \cite{val} re-observed Liller 1 in the near-IR using a standard 
ground-based filter set where a transformation to the 2MASS system is well understood. 
They found a distance modulus of 14.48 mag (d=7.87 kpc) with an uncertainty of 0.2~mag 
(0.4~kpc), in the center of the range of values for the distance from \cite{Ortolani2007}.

The results for the neutron star mass and radius determined for the distance ranges 5.8--6.5~kpc 
(Stephens et al. 2000 calibration), 7.9($\pm$0.4) kpc \citep{val}, and 8--10 kpc (other calibrations) are 
shown in Figure \ref{fig:mr}. 
The mass and radius obtained for the 
most recent distance determination by \cite{val} are consistent with the values 
found taking into account the
whole uncertainty range from \cite{Ortolani2007}.

The hydrogen mass fraction X plays a role in the Eddington limit for the luminosity (equation 1).
To obtain the results shown in Figure \ref{fig:mr} we have integrated the 
PDF for all possible values of X between zero and solar (X=0.7)
assuming solar matter is accreted, and then metallicity subsequently increased due to the nucleosynthesis 
during the thermonuclear X-ray burst. To test the sensitivity of the results on the mass-radius
plane, we have calculated the probability distribution function for two fixed, extreme values of X, 
X=0 (no hydrogen) and X=1 (pure hydrogen, even if this is an unrealistic case). Results are shown in 
Figure \ref{fig:x}. A high hydrogen mass fraction corresponds to higher masses and radii. 
Nevertheless, the fact that the hydrogen fraction plays its role in the determination of the 
Eddington limit is reflected in a higher sensitivity of the final mass to the value of X.
Since all possible values for the hydrogen mass fraction predict physically plausible 
results for the mass and radius, this test provides no constraints on X for the Rapid Burster.

Finally, the result of a similar test on the color correction factor is
also shown in Figure \ref{fig:x}. In this case, a higher color correction factor
corresponds to higher mass and radius. Since the color
correction factor plays a role in the determination of the neutron star radius 
from the emitting area (equation 2), the radius is the parameter more affected. 
Note that works on atmosphere models \citep{madej,majczyna, suleimanov} calculate 
the color correction factors only for metallicities of solar values or smaller. 
The smallest values for the color correction factor are obtained for solar abundances, i.e, the
highest metallicity considered in those works. The small color correction factor is 
due to the absorption edges of Fe, which move the neutron star hardened spectrum back towards lower energies, closer
to the blackbody. 
However, in Eddington limited bursts,
the photosphere of the neutron star is expected to show ashes from H/He burning,
produced deep in the envelope (Weinberg, Bildsten \& Schatz 2006, in 't Zand \& Weinberg 2010).
This is in agreement with results from hydrodynamic simulations, which show that convective
mixing reaches the surface only when the flux reaches the Eddington limit
(Woosley et al. 2006, Fisker et al. 2008, Jos\'e et al. 2010).
This would imply more absorption features in the final emerging 
spectrum and a smaller color correction factor,
thus our resulting PDF would move down to smaller masses and radii.
So we can consider that the color correction factors taken into account in the
present work provide, in the worst case, an upper limit for the mass and radius of 
the Rapid Burster.

\section{Summary}

The Rapid Burster showed a type I X-ray burst with photospheric radius expansion on 2009 March 5.
The analysis of Swift/XRT time-resolved spectra of that burst allows us to constrain 
the mass and radius of the neutron star in the Rapid Burster. 
After marginalization over the whole uncertainty range for the distance (5.8--10~kpc),
we obtain M=1.1$\pm$0.3~M$_{\odot}$ and R=9.6$\pm$1.5~km ($1\sigma$ uncertainties),
compatible with both normal matter and strange quark matter equations of state for the accreting object in the Rapid Burster.
We check the sensitivity of the results to the distance, the color correction factor, 
and the hydrogen mass fraction, and we find that only the distance plays a role.
Most recent distance determinations locate the Globular Cluster Liller 1 hosting
the Rapid Burster at the center of the whole uncertainty range, confirming the results for the
mass and radius found above.

As far as we know, the type I burst shown in the present work is the sole 
example from the Rapid Burster showing photospheric radius expansion. 
Therefore, future X-ray observations of the Rapid Burster 
detecting new type I bursts with photospheric expansion would provide a confirmation
of the results presented.
We have checked that neither the uncertainties in the hydrogen mass fraction nor in
the color correction factor can change our conclusions. Only the distance is the 
key factor in this case, and thus a better determination of the mass and radius
of the neutron star in the Rapid Burster requires a 
definitive measurement of the distance to Liller 1.


\acknowledgments
We thank Wolfgang Pietsch and Paula Coelho 
for useful and helpful suggestions and discussions. 
We acknowledge the Swift PI, Niel Gehrels, and the science and mission operations 
teams for their support of these TOO observations. 
This research has been funded by the MICINN grants AYA2010-15685 and AYA2011-23102, and the 
ESF EUROCORES Program EuroGENESIS through the MICINN grant EUI2009-04167.

\end{document}